\documentclass[12pt,preprint]{aastex}

\usepackage{epsfig}

\def\vkm{km s$^{-1}$}

\def\degree{$^\circ$}
\def\arcs#1{$#1''$}
\def\arcsa#1#2{$#1^{\prime\prime}_{^\textrm{.}}#2$}

\def\smassrate{$M_\odot$ yr$^{-1}$}
\def\solarmass{$M_\odot$}

\def\Jyb{Jy beam$^{-1}$}
\def\mJyb{mJy beam$^{-1}$}
\def\Jybk{Jy beam$^{-1}$ km s$^{-1}$}
\def\Tk{K km s$^{-1}$}
\def\tlabel#1{(\textit{#1})}

\def\cmc{cm$^{-3}$}
\def\cms{cm$^{-2}$}
\def\micron{$\mu$m}
                                                                                
\def\ra#1#2#3#4{#1^\mathrm{h} #2^\mathrm{m} #3^\mathrm{s}_{^\textrm{.}} #4}
\def\dec#1#2#3#4{#1\degr #2\arcmin #3^{\prime\prime}_{^\textrm{.}}#4}

\def\Lbol{L_\textrm{\scriptsize bol}}

\def\H2{H$_2$}
\def\N2HP{N$_2$H$^+$}

\def\NH3{NH$_3$}

\def\aHCOP{H$^{13}$CO$^+$}
                                                                                
\def\putfig#1#2#3{\epsfig{scale=#1,angle=#2,figure=#3}}
\def\leftblank#1{}

\begin{document}

\title{Submillimeter arcsecond-resolution mapping of the highly collimated protostellar jet HH 211}
\author{Chin-Fei Lee\altaffilmark{1,2}, 
Paul T.P. Ho\altaffilmark{1,3}, Aina Palau\altaffilmark{4},
Naomi Hirano\altaffilmark{1}, 
Tyler L. Bourke\altaffilmark{3}, 
Hsien Shang\altaffilmark{1}, and Qizhou Zhang\altaffilmark{3}
}
\altaffiltext{1}{Academia Sinica Institute of Astronomy and Astrophysics,
P.O. Box 23-141, Taipei 106, Taiwan; cflee@asiaa.sinica.edu.tw}
\altaffiltext{2}{
Harvard-Smithsonian Center for Astrophysics, Submillimeter Array,
645 North A'ohoku, Hilo, HI 96720}
\altaffiltext{3}{Harvard-Smithsonian Center for Astrophysics, 60 Garden
Street, Cambridge, MA 02138}
\altaffiltext{4}{Laboratorio de Astrof\'{\i}sica Espacial y F\'{\i}sica
Fundamental, INTA, Apartado 50727, E-28080 Madrid, Spain}

\begin{abstract}
We have mapped the protostellar jet HH 211 in 342 GHz continuum, SiO
($J=8-7$), and CO ($J=3-2$) emission at $\sim$ \arcs{1} resolution 
with the Submillimeter Array (SMA).  
Thermal dust emission is seen in continuum at the center of the jet, 
tracing an envelope and a possible optically thick compact disk 
(with a size $<$ 130 AU) around the protostar.
A knotty jet is seen in CO and SiO as in \H2{},
but extending closer to the protostar.
It consists of a chain of knots
on each side of the protostar, with an interknot spacing of $\sim$
\arcs{2}$-$\arcs{3} or 600$-$900 AU and
the innermost pair of knots at only $\sim$ \arcsa{1}{7} or 535 AU from the
protostar. These knots
likely trace unresolved internal (bow) shocks (i.e., working surfaces)
in the jet, with a velocity range up to $\sim$ 25 \vkm{}.
The two-sided mass-loss rate of the jet is estimated to be 
$\sim (0.7-2.8)\times 10^{-6}$ \solarmass{} yr$^{-1}$. 
The jet is episodic, precessing, and bending.
A velocity gradient is seen consistently across 
two bright SiO knots (BK3 and RK2)
perpendicular to the jet axis, 
with $\sim$ 1.5$\pm$0.8 \vkm{} at $\sim$ 30$\pm$15 AU, suggesting a presence
of a jet rotation. 
The launching radius of the jet, derived
from the potential jet rotation,
is $\sim$ 0.15$-$0.06 AU in the inner disk.
\end{abstract}


\section{Introduction}

Protostellar jets are seen associated with
low-mass protostars in the early stages of star formation.
In spite of numerous studies,
their physical properties (e.g., speed, episodic nature, 
collimation, and angular momentum) and thus launching mechanisms
are still not well understood.
They are believed to be launched from accretion disks around the protostars
\cite[see recent reviews by e.g.,][]{Pudritz2007,Ray2007,Shang2007},
allowing us to probe the accretion process, 
which remains heretofore unresolved, as it requires us to observe
directly in the inner parts at the AU scale.
The Submillimeter Array (SMA)\footnote{ The Submillimeter Array is a joint project between the
Smithsonian Astrophysical Observatory and the Academia Sinica Institute of
Astronomy and Astrophysics, and is funded by the Smithsonian Institution and
the Academia Sinica.} \citep{Ho2004}, 
with the capability to probe warm and dense molecular gas at high angular
resolution, can be and has been used to
study the physical properties of the jets in detail
\cite[e.g.,][]{Hirano2006,Lee2007}.

The HH 211 outflow is an archetypical outflow with a highly collimated jet,
located in the IC 348 complex in Perseus.
The distance of the IC 348 complex is assumed to
be 320 pc \citep{Lada2006}, but it could be 250 pc
\citep{Enoch2006}.
The outflow was discovered in \H2{} shock emission at 2.12 \micron{}
\citep{McCaughrean1994}, powered by a young, low-mass, and low-luminosity
($\sim 3.6\; L_\odot$) Class 0 protostar with $T_{bol} < 33$ K
\citep{Froebrich2005}. A collimated CO jet was seen surrounded by cavity
walls in CO (J=2-1) (Gueth \&  Guilloteau 1999, hereafter
\citet{Gueth1999}).
Recent observations with the SMA in SiO (J=5-4) \citep{Hirano2006} and
(J=8-7) \citep{Palau2006} also revealed a collimated SiO
jet consisting of a chain of spatially unresolved knots aligned with the CO jet, tracing shock
emission along the jet. In this paper, we present observations of the jet in
SiO (J=8-7) and CO (J=3-2) at higher-angular resolution than in
\citet{Palau2006}, in order to better
constrain the physical properties and thus the launching mechanisms 
of the jet. 

\section{Observations}\label{sec:obs}

Observations toward the HH 211 jet were carried out with the SMA
on 2004 October 4 and 18 in the compact
configuration and on 2004 September 10 in the extended configuration.
Note that
the observation on 2004 October 18 was only a short 
partial track 
and thus not included in our analysis.
The zenith opacities were $\tau_{230}\simeq 0.1$ and 0.13,
respectively, on 2004 October 4 and 2004 September 10.
SiO ($J=8-7$) and CO ($J=3-2$) lines were observed simultaneously with
continuum using the 345 GHz band receivers. 
The receivers have two sidebands, lower and upper,
covering the frequency range from 335.58 to 337.55 and from 345.59 to 347.56 GHz,
respectively. 
Combining the line-free portions of the two sidebands results in
a total continuum bandwidth of $\sim$ 3.7 GHz centered
at $\sim$ 342 GHz (or $\lambda \sim$ 880 \micron{}) for the continuum.
The baselines have projected lengths ranging from $\sim$ 20 to 225 m.
The primary beam has a size of $\sim$ \arcs{35} and three pointings
were used to map the jet.
For the correlator,
128 spectral channels were used for each 104 MHz chunk,
resulting in a velocity resolution of $\sim$ 0.7 \vkm{} per channel.

The visibility data were calibrated with the MIR package,
with Saturn, Venus, quasars 3C84 and J0355+508 as passband calibrators, 
quasars 3C84 and J0355+508 as gain calibrators, and Uranus ($\sim 70$ Jy)
as a flux calibrator.
The flux uncertainty is estimated to be $\sim$ 20\%.
The calibrated visibility data were imaged with the MIRIAD package.
The dirty maps that were produced from the calibrated visibility data
were CLEANed using the Steer clean method,
producing the CLEAN component maps.
The final maps were obtained by restoring the CLEAN component
maps with a synthesized 
(Gaussian) beam fitted to the main lobe of the dirty beam. 
With natural weighting, the synthesized beam has 
a size of \arcsa{1}{28}$\times$\arcsa{0}{84} 
at a position angle (P.A.) of $\sim$ 70\degree{}.
The rms noise level is $\sim$ 0.33 \Jyb{} in the 
channel maps and 6.5 \mJyb{} in the continuum map.
The velocities of the channel maps are LSR.
The typical rms of the gain phases is $\sim$ 20\degree{},
resulting in an absolute positional accuracy of
one tenth of the synthesized beam, or $\sim$ \arcsa{0}{1}.

\section{Results}

Our results are presented in comparison to
the IR image (\H2{} at 2.12 \micron{} + continuum)
made with the VLT on 2002 January 4 \citep{Hirano2006}, which
shows clear shock interactions along the jet axis.
Since our observations were carried out $\sim$ 3 years later than
the IR image, the \H2{} shock knots and bow shocks might have
moved down along the jet axis by $\sim$ \arcsa{0}{2}$-$\arcsa{0}{4}
from their positions in that image, assuming a typical jet velocity of 
100$-$200 \vkm{}.
This movement, however, does not affect significantly our comparison and conclusions,
considering that the angular resolution of our observations is $\sim$
1\arcsec{} along the jet axis.
The systemic velocity in this region is assumed to be
$9.2$ \vkm{} LSR, as in \citet{Hirano2006} and \citet{Palau2006}.
Throughout this paper, the velocity is relative to this systemic value.

\subsection{342 GHz Continuum Emission} \label{sec:cont}

\begin{figure} [!hbp]
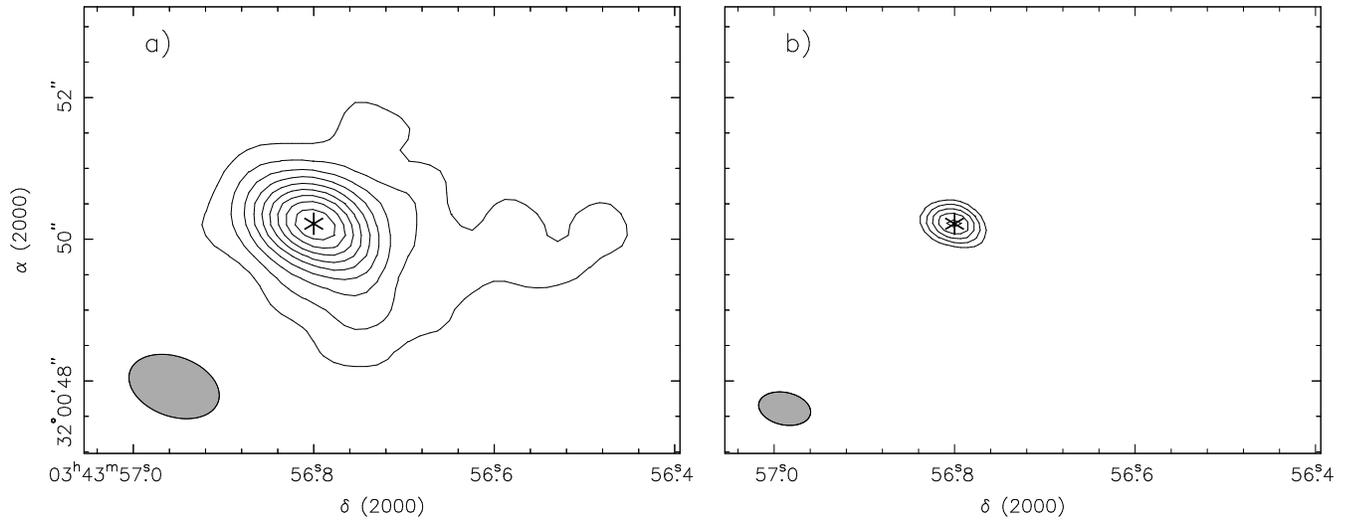

\centering
\putfig{0.75}{270}{f1.ps}
\figcaption[]
{342 GHz continuum maps with the asterisk marking the source position.
\tlabel{a} Map made using all the available visibility data. The contours
go from 10 to 90\% of the peak value, which is 155 \mJyb{}. 
The beam is \arcsa{1}{28}$\times$\arcsa{0}{84} 
with a P.A. of $\sim$ 70\degree{}.
\tlabel{b} Map made using the visibility data with the $uv$ distance greater
than 100 k$\lambda$, showing the central compact source. The contours
go from 35 to 95\% with a step of 15\% of the peak value, which is
77 \mJyb{}. The beam is \arcsa{0}{74}$\times$\arcsa{0}{46}
with a P.A. of $\sim$ 78\degree{}.
\label{fig:cont}}
\end{figure}

Continuum emission is detected at 342 GHz
at the center of the \H2{} flow 
with a total (integrated) flux of 0.44$\pm0.10$ Jy.
It has a peak at
$\alpha_{(2000)}=\ra{03}{43}{56}{801}$,
$\delta_{(2000)}=\dec{32}{00}{50}{22}$, with a positional
uncertainty of \arcsa{0}{1} (Fig. \ref{fig:cont}).
This peak position is within \arcsa{0}{1} of that
found at 43.3 GHz (or $\lambda= 7$ mm) with the VLA at an
angular resolution of $\sim$ \arcsa{0}{15} \citep{Avila2001}, and is thus
considered as the position of the protostar throughout this paper.
Note that the continuum flux here is about twice as that found in
\citet{Palau2006}, which was based on
the observation on 2004 October 18.
As mentioned, that observation was only a short partial track with 
poor $uv$ coverage.
Thus, even
though their visibility amplitude versus $uv$ distance plot
is similar to ours,
most of the extended emission was not recovered in their image 
(see their Fig. 1b).

\begin{figure} [!hbp]
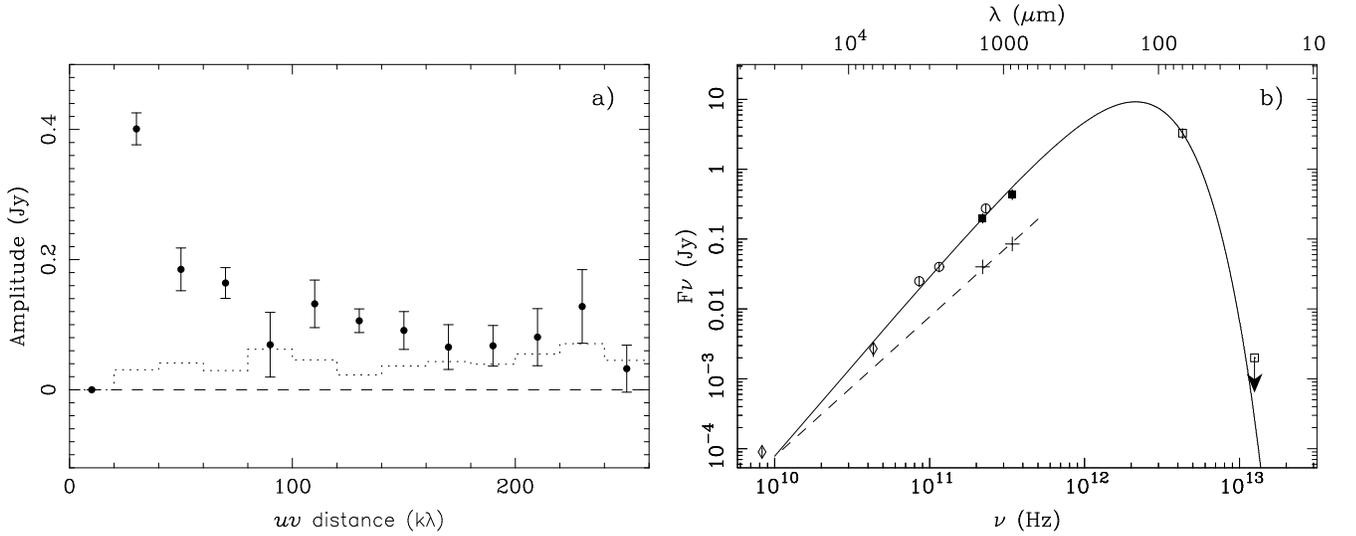

\centering
\putfig{0.7}{270}{f2.ps}
\figcaption[]
{\tlabel{a} Visibility amplitude versus $uv$ distance plot 
for the continuum source with 1 $\sigma$ error bars.
The dotted histogram is the zero-expectation level ($\sim$ 1.25 $\sigma$).
\tlabel{b} Spectral energy distribution (SED)
of the continuum source. The filled squares are from our SMA observations.
The open squares, open circles, and diamonds are from the Spitzer [70
\micron{} from \citet{Rebull2007} and 24 \micron{} from Karl
Stapelfeldt, private communication], PdBI \citep{Gueth1999}, and VLA
\citep{Avila2001} observations, respectively. The crosses indicate the
fluxes from the central unresolved point source estimated from our
SMA observations at 342 and 220 GHz.
The solid line shows the single-temperature fit
to the SED of the continuum source (see text for detail).  The dashed line
shows the relation $F_{\nu} \propto \nu^2$ for the central
compact source.
\label{fig:uvsedcont}}
\end{figure}

The emission is seen extending $\sim$
\arcs{2} to the southwest from the protostar
roughly perpendicular to the jet axis,
similar to that seen at 230 GHz \citep{Gueth1999}, 
likely tracing the flattened envelope perpendicular to the jet axis.
The envelope, however, seems asymmetric with less emission
extending to the northeast.
Faint emission is also seen extending to the northwest, 
similar to that seen at 220 GHz \citep{Hirano2006},
probably tracing the envelope material around the west outflow lobe.
Near the protostar, the structure is compact and not resolved, as
seen in the map made using the visibility data with the $uv$ distance greater
than 100 k$\lambda$ (see Fig. \ref{fig:cont}b), with a flux of $\sim$
0.08$\pm0.02$ Jy.
This unresolved compact source is also seen in the 
visibility amplitude versus $uv$
distance plot, with a similar flux (Fig. \ref{fig:uvsedcont}a).
The size (diameter) of this compact source has an upper limit of
$\sim$ \arcsa{0}{4} or 130 AU set by the longest baseline. 
Two components, an envelope and a compact source, have also been
suggested in other Class 0 sources, e.g., 
IRAS 16293-2422 and L1448-C \citep{Schoier2004}, and HH 212
\citep{Codella2007,Lee2007}.

The spectral energy distribution (SED) of the continuum source
(see Fig. \ref{fig:uvsedcont}b) 
indicates that the continuum emission at 342 GHz is 
mainly thermal dust emission. 
Assuming a constant temperature $T_d$, a frequency-independent source
size $\Omega$, and a mass opacity
$\kappa_\nu = 0.1 (\nu/10^{12} \textrm{\scriptsize Hz})^\beta$
cm$^2$ g$^{-1}$ \citep{Beckwith1990} for the dust, the SED can be fitted with
$T_d \sim$ 30 K, $\Omega \sim$ 3.6 arcsec$^2$,
$\beta \sim 0.6$, and an optical depth  $\tau_\nu \sim$ 0.086 at 342 GHz.
Our value of $\beta$ is similar to that found by 
\cite{Gueth1999} and \cite{Avila2001}.
The dust temperature is also consistent with the bolometric temperature,
which was found to be $<33$ K \citep{Froebrich2005}.
However, our $\beta$ is smaller than that found
including fluxes from the larger-scale envelope (with a mass of
$\sim$ 0.8 \solarmass) in photometric broadband observations, 
which is 1.3 \citep{Froebrich2005}.
$\beta > 1$ was also found toward large-scale envelopes around other
embedded YSOs \citep{Dent1998}. 
Thus,
excluding the larger-scale envelope tends to decrease the value of $\beta$,
suggesting that the dust grains grow bigger toward the source.
The compact source at the center may have even lower $\beta$.
To examine this, fluxes of the compact source at 342 and
220 GHz are compared with $F_{\nu} \propto \nu^2$ or $\beta=0$. 
Assuming that fluxes at long baselines are dominated by the compact source,
the fluxes are $\sim$ 0.08$\pm0.02$ Jy at 342 GHz 
and 0.04$\pm0.02$ Jy at 220 GHz
\cite[3 $\sigma$ detection with the SMA data in][with 1 $\sigma$
$\sim$ 0.013 \Jyb{} and the uncertainty in flux scale $\sim$ 20 \%]{Hirano2006}.
The SED of the compact source at those frequencies seems roughly 
consistent with $\beta=0$, or an optically thick emission, as in HH 212 
\citep{Codella2007}.
In addition, half of the emission at
43.3 GHz could be from the unresolved compact source.
If the compact emission at 342 and
43.3 GHz are indeed arisen from the same region, then the
compact source has a size (diameter) of $\lesssim$ \arcsa{0}{1} or 30 AU
\citep{Avila2001}.
Therefore, the compact source probably has
different origin and is likely to be a warm 
(with a brightness temperature $> 80$ K) optically thick (accretion) disk
deeply embedded in the cold envelope.
Observations at higher angular resolution are really needed to
confirm this.

The total mass of the continuum emission 
(including both the extended and compact components)
is estimated to be $\sim$ 0.05 \solarmass{}, assuming
optically thin emission with a temperature of 30 K.
Since part of the emission is likely to be optically thick, 
the mass here is only a lower limit. 

\subsection{SiO Jet}

A knotty jet is seen in SiO, as in \citet{Palau2006}, but with
more knots resolved at higher angular resolution (Fig. \ref{fig:jet}c).
It is bipolar with the blueshifted side in the southeast and 
the redshifted side in the northwest, but with the redshifted side
brighter than the blueshifted side.
It consists of a chain of knots
on each side of the source with an interknot spacing of $\sim$
\arcs{2}$-$\arcs{3} or 600$-$900 AU.
It extends out to $\sim$ \arcs{18} away, with
the innermost pair of knots at only $\sim$ \arcsa{1}{7} or 535 AU from the
source.
Most of the knots have \H2{} counterparts, except for those in
the inner part. The lack of \H2{} counterparts there was already seen by
\citet{Hirano2006} and attributed to
the heavy dust extinction associated with the dense envelope
as seen in \aHCOP{} \citep{Gueth1999} and NH$_3$ \citep{Wiseman2001}.

\subsection{Jet Axes}\label{sec:jetaxis}

\begin{figure} [!hbp]
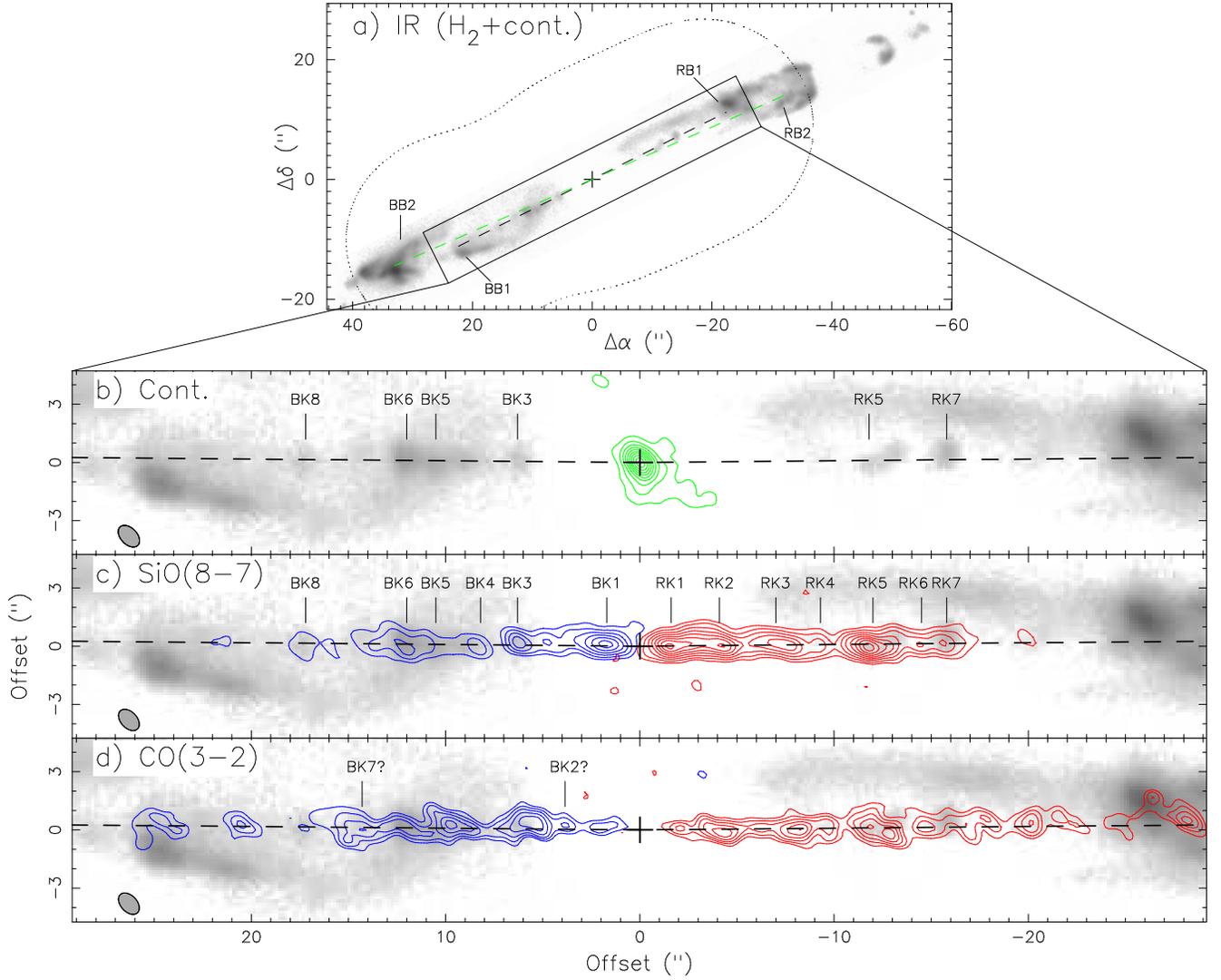

\centering
\putfig{0.85}{270}{f3.ps}
\figcaption[]
{ 
\tlabel{a} The IR image adopted from \citet{Hirano2006}.
The dotted line outlines our observed region.
The dashed lines indicate the jet axes.
The cross marks the source position. Here BBx and RBx indicate blueshifted and
redshifted bow shocks, respectively.
\tlabel{b} 342 GHz continuum contours (as in Fig. \ref{fig:cont}a)
on top of the IR image.
\tlabel{c} Redshifted (6 to 34 \vkm{}) and 
blueshifted (-22 to -2 \vkm{}) SiO 
contours on top of the IR image.
Contour spacing is 4 \Jybk{} (37.8 \Tk) with the first contour at 4 \Jybk{}.
Here, knots RK1, RK3, RK5, RK7, BK1, BK3, BK5, BK6, and BK8
correspond to knots R1, R2, R3, R4, B1, B2, B3, B4, and B5, respectively, 
in \citet{Hirano2006}.
Here BKx and RKx indicate blueshifted and redshifted knots, respectively.
\tlabel{d} High-velocity redshifted (10.3 to 35.3 \vkm{}) and blueshifted
(-23.7 to -8.7 \vkm{}) CO 
contours on top of the IR image.
Contour spacing is 1.5 \Jybk{} (14.2 \Tk) 
with the first contour at 3 \Jybk{} (28.4 \Tk).
In \tlabel{b}, \tlabel{c}, and \tlabel{d},
the beams are \arcsa{1}{28}$\times$\arcsa{0}{84} and 
the images are rotated by 26.6\degree{} clockwise. The western and eastern
components of the jet axis are seen bent by $\sim$ 0.5\degree{} to the north.
\label{fig:jet}}
\end{figure}

The eastern component and western component of the SiO jet are not exactly
antiparallel (Fig. \ref{fig:jet}c). They
are misaligned slightly by $\sim$ 1\degree{}, with their axes estimated to
have a P.A. of 116.1\degree{}$\pm0.5$\degree{} and
297.1\degree{}$\pm0.5$\degree{}, respectively, by connecting the source to
the SiO knots. 
Since their original paths of motion are likely to be antiparallel, this
misalignment suggests a presence of a jet bending.
Assuming both bent by the same degree, they
are both bent by $\sim$ 0.5\degree{} to the north 
with a jet (or mean) axis having a P.A. of 116.6\degree{}$\pm0.5$\degree{}.
The peaks of the closest pair of the SiO knots (BK1 and RK1) are not exactly
aligned with this jet axis, probably suggesting a slight precession of the jet.
That the jet is slightly sinuous also supports this possibility.
The angle of precession is estimated to be $<$ 1\degree{}.

The jet may also have a large-scale precession as discussed in
\citet{Gueth1999} and \citet{Eisloffel2003}, with an angle of $\sim
3$\degree{}. The axis of the SiO jet, which
is aligned with the \H2{} knots, is considered as the jet axis in the inner
part out to \H2{} bow shocks BB1 and RB1 (Fig. \ref{fig:jet}a). The jet axis
in the outer part, which can be estimated by connecting the source to the
tip of bow shock BB2 further out in the east, is found to be different
by $\sim$ 3\degree{} with a P.A. $\sim$ 113.6\degree{} (indicated by the
green dashed line). If we connect this axis to the west, the \H2{}
emission RB2 could be the counterpart of bow shock BB2 but mainly seen with
the southern wing. 

\subsection{CO Jet and Shells} \label{sec:COjs}

CO emission is detected not only along the jet axis but also toward outflow
shells. In the following, two velocity ranges,
high (from $-$23.7 to $-$8.7 \vkm{} and from 10.3 to 35.3 \vkm{})  and
low  (from $-$5.7 to $-$1.7 \vkm{} and from 0.3 to 4.3 \vkm{},
respectively for the blueshifted emission and redshifted emission),
are selected to show these components.

A knotty jet-like structure is also seen in CO here
at high velocity
(Fig. \ref{fig:jet}d), as in lower excitation line 
\citep{Gueth1999}. However, this jet-like
structure actually contains two components, a jet component
with a similar velocity structure to the SiO jet and a
high-velocity shell component with a different velocity structure
(see next section).
With the selected velocity range, 
the high-velocity CO emission at $\sim$ \arcs{2} 
away from the source is mainly from the jet component. It is faint, 
associated with knots BK1 and RK1 seen in SiO, but slightly
downstream.

\begin{figure} [!hbp]
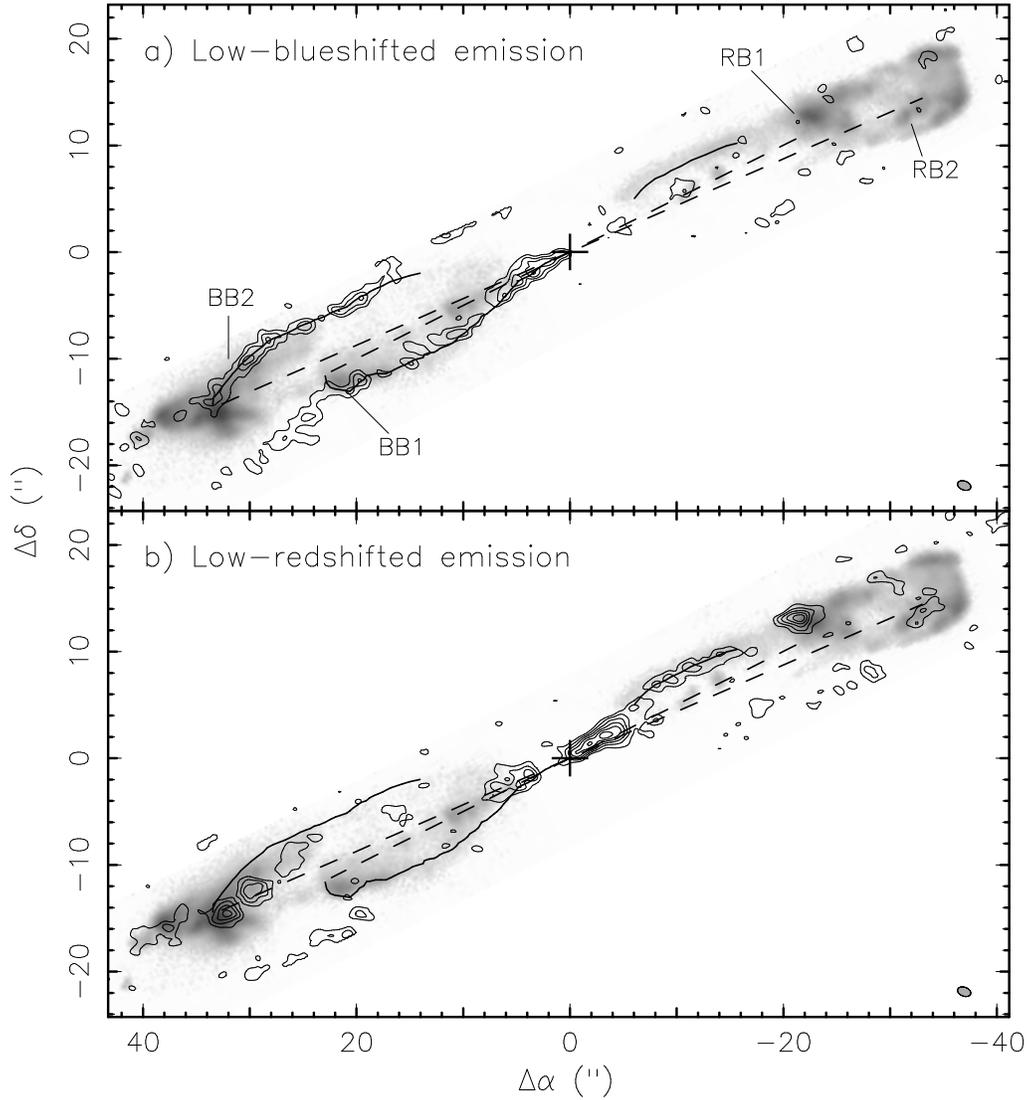

\centering
\putfig{0.85}{270}{f4.ps}
\figcaption[]
{Low-velocity CO contours on top of the IR image.
The beams are \arcsa{1}{28}$\times$\arcsa{0}{84}.
The cross marks the source position.
\tlabel{a} shows the low-redshifted CO emission integrated from 0.3 to 4.3 
\vkm{}.
\tlabel{b} shows the low-blueshifted CO emission integrated
from  $-$5.7 to $-$1.7 \vkm{}.
Contour spacing is 1.4 \Jybk{} with the first contour at 1.4 \Jybk{}.
\label{fig:COshell}
}
\end{figure}

At low velocity, rim-brightened shell-like structures
are seen in CO (Fig. \ref{fig:COshell}), 
coincident with those seen in the IR image, which 
are mainly from the continuum except near the bow shocks 
where the \H2{} emission dominates \citep{Eisloffel2003}.
These shell-like structures were also seen in lower
excitation lines of CO
and thought to trace the dense cavity walls of the outflow lobes
produced by the \H2{} bow shocks located at the ends of the jet
\citep{Gueth1999}.
In the eastern lobe, however, the southern and northern
shells are seen associated with two different bow shocks,
BB1 and BB2, respectively. Thus,
the shells likely trace the dense cavity walls
recently shocked (excited) by the internal \H2{} bow shocks
at different jet axes
because of a large-scale jet precession (see \S \ref{sec:jetaxis}).

\subsection{Kinematics}\label{sec:kinematics}

\begin{figure} [!hbp]
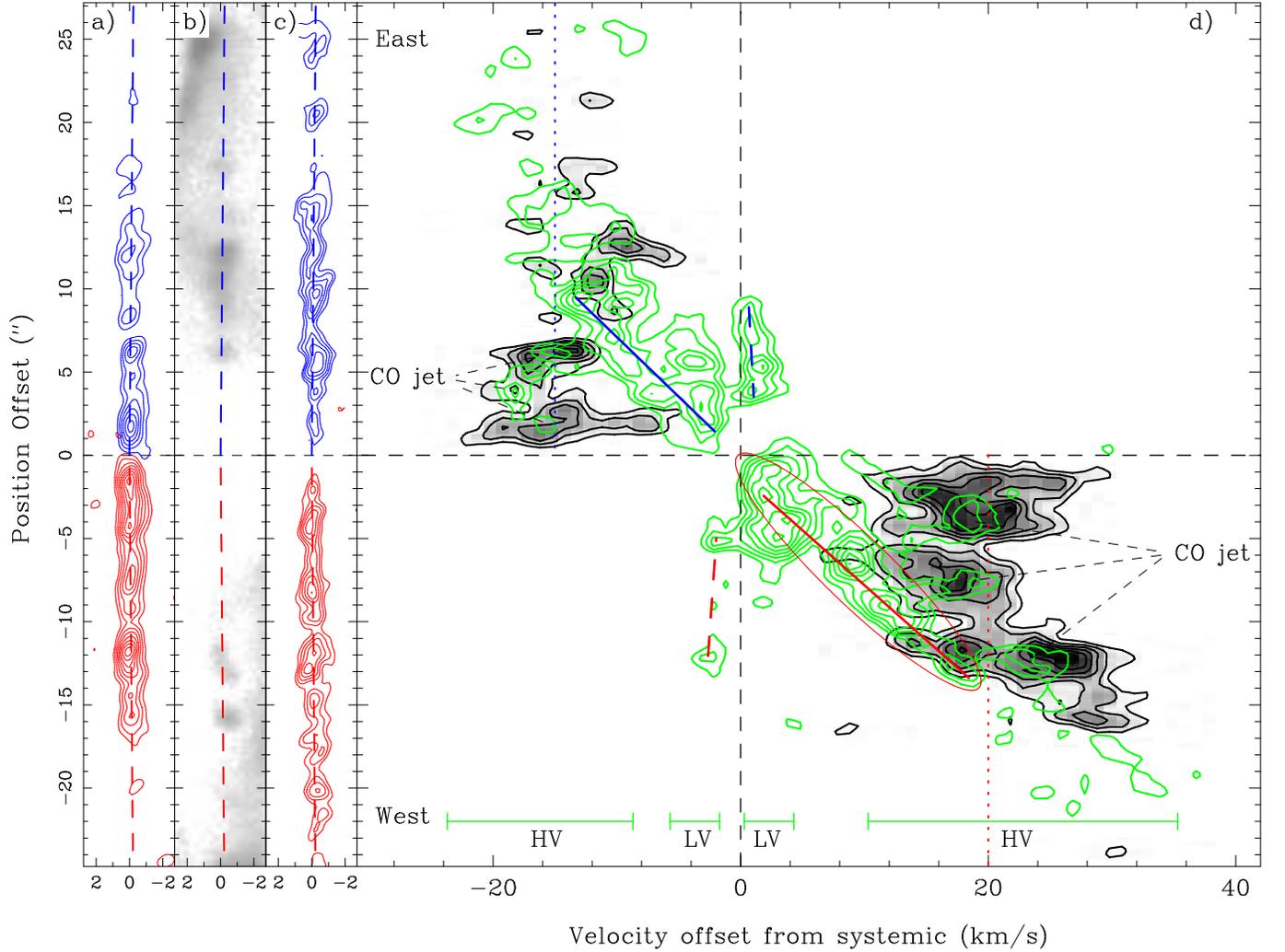

\centering
\putfig{0.7}{270}{f5.ps}
\figcaption[]
{PV diagrams of the SiO and CO emission cut along the jet axis
with a width of \arcs{1}.
\tlabel{a} SiO contours, \tlabel{b} IR image, and \tlabel{c} high-velocity
CO contours, as in Fig. \ref{fig:jet}.
\tlabel{d} PV diagrams of the SiO (black contours with image) and CO 
(green contours) emission. The SiO contours have a spacing of 0.25 \Jyb{}
(2.36 K) with the first contour at 0.5 \Jyb{} (4.73 K). 
The CO contours have a 
spacing of 0.225 \Jyb{} (2.12 K)  with the first contour at 0.45 \Jyb{}
(4.25 K).
The red and blue dashed lines mark the emission from the low-velocity shells.
The red and blue solid lines mark the emission from the high-velocity shells.
The ellipse outlines the emission from the high-velocity shell in the west.
The red and blue dotted lines indicate the mean velocities of the jet
on the redshifted and blueshifted sides, respectively.
Here, HV and LV denote the high-velocity and
low-velocity ranges, respectively, for the CO emission. 
\label{fig:pvjet}
}
\end{figure}

Position-velocity (PV) diagrams of the SiO and CO emission 
cut along the jet axis are used to study
the kinematics of the jet (Fig. \ref{fig:pvjet}).
Note that the CO emission seen at low velocity
(indicated by dashed lines) likely traces the
cavity walls along the jet axis.

\begin{figure} [!hbp]
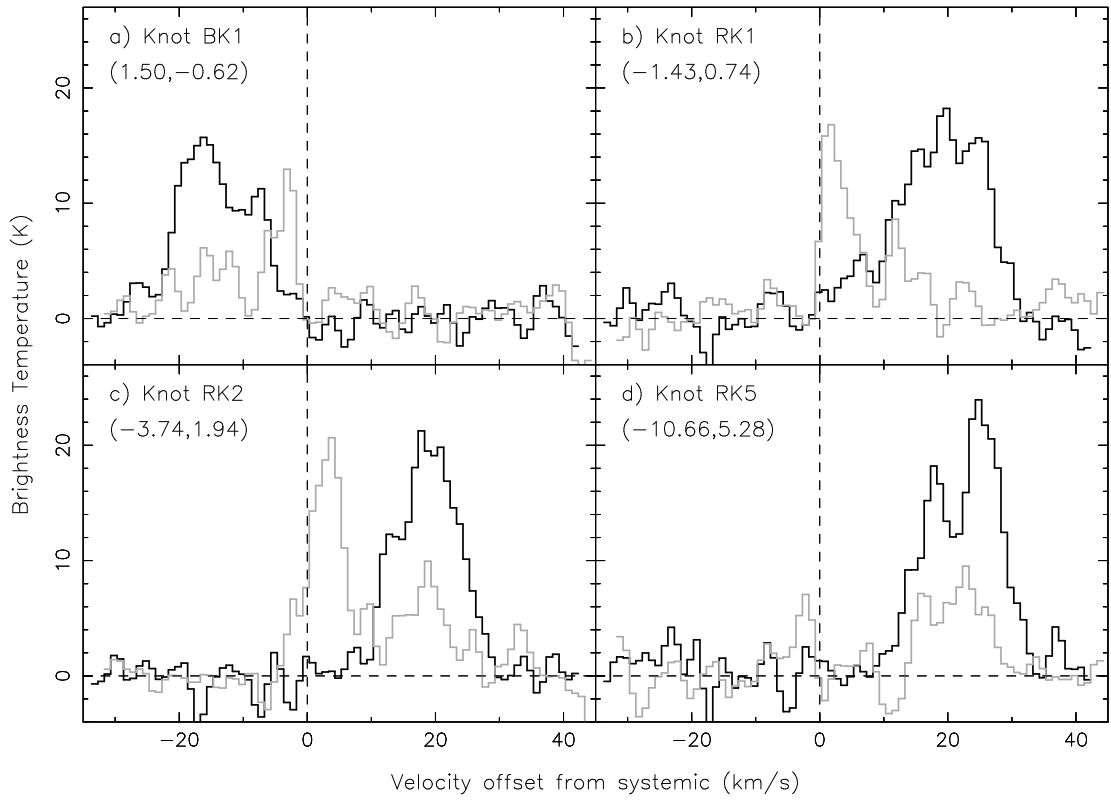

\centering
\putfig{0.6}{270}{f6.ps}
\figcaption[]
{SiO (dark) and CO (gray) spectra toward four bright knots with their positions
given in the upper left corners. 
\label{fig:specSiO}
}
\end{figure}

In the jet, the SiO emission is localized toward the knots, with 
a range of velocities detached from the systemic velocity:
knots BK1, RK1, RK2, and RK5 are bright
and seen with a velocity range of $\sim$ 25
\vkm{} (see also their spectra in Fig. \ref{fig:specSiO}),
while other knots are faint and seen with a narrower velocity range. At
high velocity, CO emission (labeled CO jet)
also arises from the jet, showing
a similar velocity structure with similar mean velocity
but with narrower velocity range (see also  Fig. \ref{fig:specSiO}).
Thus, the SiO emission 
and the CO jet emission can be used together to study the kinematics of the jet.
They are blueshifted in the east and
redshifted in the west. They are,
however, more redshifted (with a
mean velocity of $\sim$ $20$ \vkm{})
in the west than blueshifted (with a
mean velocity of $\sim$ $-15$ \vkm{})
in the east. In addition, for each component of the jet,
the velocity centroids of the individual knots are also
different. 

Linear velocity structures (indicated with solid lines) are
seen in CO, as in lower excitation line \citep{Gueth1999},
with the velocity magnitude increasing with the distance from the source.
These velocity structures are different from that of the SiO jet.
To study their origin,
we plot in Figure \ref{fig:shell2} the
emission of the linear velocity structure (marked by an ellipse) in the west, 
where it can be better separated from the jet emission.
The emission is seen around the jet axis but brighter in the north, with 
the transverse width increasing with the distance from the source,
surrounded by the rim-brightened IR shell (cavity wall).
In the far end, the emission shows shell-like structures associated with the
\H2{} knots. In the near end, however, it is unclear but may
show shell-like structures associated with the SiO knots (comparing Fig.
\ref{fig:shell2} with Fig. \ref{fig:jet}c).
PV cuts across the jet axis in the west
at \arcs{11} and \arcs{12} in the far end, where the structure is
better resolved, also show a ring-like velocity structure as expected for a 
shell (Fig. \ref{fig:pv_per}). Note that, in these PV cuts, 
the CO emission seen at low velocity is likely from the cavity wall.
Therefore, the CO emission associated with the linear velocity structures
likely traces the (internal) shells or wakes driven by the (internal)
bow shocks \cite[see, e.g.,][]{Raga1993,Gueth1999,Lee2001},
surrounded by the dense cavity walls.

\subsection{Temperature, Column Density, and Density}\label{sec:TCD}

The CO emission is assumed to have
an excitation temperature of 100 K in the jet and
50 K in the high-velocity shells.
Note that, however, the CO emission, 
which was detected also in higher-$J$ transitions in the far infrared, 
could have a higher excitation temperature \citep{Giannini2001}.
The \H2{} column density can be derived assuming a CO
abundance of 8.5$\times10^{-5}$ \citep{Frerking1982} and optically thin
emission in local thermal equilibrium. 
It is found to be $(4-8)\times 10^{20}$ \cms{} in the jet,
which has an intensity of 57$-$104 \Tk{} toward the knots 
(see Fig. \ref{fig:jet}d, integrated from -23.7 to -8.7 \vkm{} and 10.3 to 35.3
\vkm{}),
and $4.5\times 10^{20}$ \cms{} in the high-velocity shells, which have
a mean intensity of 85 \Tk
(see Fig. \ref{fig:shell2}, integrated from 0.3 to 20.3 \vkm{}).
The jet is unresolved with a size (diameter) of $<$ \arcs{1} and thus
has a density of $>$ $10^{5}$ \cmc{}. 
Assuming a shell thickness of $\sim$ \arcs{1}, the high-velocity shells
have a density of $\sim$ $9\times10^4$ \cmc{}.
Note that, however, it is not meaningful to derive the density for the low-velocity
shells because their fluxes are mostly resolved out in our observations
\cite[see also][]{Gueth1999}.

Using the SiO($J=5-4$) observations from \citet{Hirano2006},
the line ratio of SiO($J=8-7$)/SiO($J=5-4$) is found to be 
$\sim$ 0.7 for the first pair of knots but decreasing to 
$\sim$ 0.4 at the far ends of the SiO jet (Fig. \ref{fig:SiOratio}).
Note that this line ratio has been found to be $\sim$ 1.0 for the first
pair of knots but decreasing to $\sim$ 0.5 at the far ends of the SiO jet by
\citet{Palau2006} at lower angular resolution. However,
with their maps, we
recalculated the line ratio at lower angular resolution using the same
velocity interval and intensity cutoff level and found it to be $\sim$ 0.8 
for the first pair of knots, and thus consistent with ours within the flux uncertainty.
The kinetic temperature of the SiO emission can be assumed
to be 300 K for the first pair of knots but 100 K for the knots further out
\citep{Hirano2006}. The density can be estimated by comparing the line ratio
to that in the LVG calculations with an assumption of optically thin
emission \citep{Nisini2002}. Note that, however, the emission could be
optically thick \citep{Cabrit2007}. The density is estimated to be
$\sim 4\times10^6$ \cmc{} for the first pair of knots but decreasing to
$\sim 3\times10^6$ \cmc{} at the far ends of the SiO jet.
These densities, however, are a factor of $\sim$ 2 lower than those
derived from the line ratio of
SiO($J=5-4$)/SiO($J=1-0$) in \citet{Hirano2006}.
A detailed study with multiple transitions of SiO will address this in a
later paper (Hirano et al. in prep).

\section{Molecular Jet}

\subsection{Shocks}

The SiO knots are seen with a range of velocities, likely
tracing the unresolved internal (bow) shocks (i.e., working surfaces)
in the jet, as in HH 212 \citep{Codella2007,Lee2007}.
It is thought that SiO abundance is greatly enhanced in the shocks
as a consequence of grain sputtering or grain-grain collisions releasing
Si-bearing material into the gas phase, which reacts rapidly with O-bearing
species (e.g., O$_2$ and OH) to form SiO
\citep{Schilke1997,Caselli1997}.
The emission is consistent with its production in C-type shocks, with a
velocity range of $\sim$ 25 \vkm{}, similar to that found to produce
the observed SiO column densities in molecular outflows \citep{Schilke1997}.
At high velocity, CO is also seen tracing the shocks in the jet. It has
smaller velocity range than SiO, tracing weaker shocks where the
temperature and density are both lower. Observations at higher angular
resolution are needed to study the morphological relationship 
between the CO and SiO emission by resolving their structures.

\subsection{Inclination}\label{sec:dis_inc}

The jet inclination, $i$, can be estimated from the mean velocity of the SiO
emission, $v_m$, with $i = \sin ^{-1} (v_m/v_j)$, where $v_j$ is the jet
velocity. 
The eastern and western components of the jet
may have different inclinations because of their different mean velocities.
Assuming a typical jet velocity of 100$-$200 \vkm{},
the eastern and western components of the jet have an inclination of
$-$8.6\degree{} to $-$4.3\degree{} and
11.4\degree{} to 5.7\degree{}, respectively. Note that, however, it is also
possible that the eastern and western components of the jet have the same
inclinations but different jet velocities as in T Tauri star jets
\citep{LopezM2003}.

\subsection{Mass-loss rate}

The mass-loss rate of the jet can be estimated from the CO emission of the
knots. As mentioned, the knots likely trace the internal (bow) shocks or
the jet itself but highly compressed by shocks. 
Since the knots are
unresolved, the compression factor is assumed to be $\sim$ 3,
as found in HH 212 at similar angular resolution \citep{Lee2007}.
Note that, however, the actual compression factor due to the shocks
could be higher, if the knots are resolved.
Thus, the (two-sided) mass-loss rate 
is given by
\begin{equation}
\dot{M}_j \sim \frac{2}{3}v_j m_{\textrm{\scriptsize H}_2} N b
\end{equation}
where $N$ and $b$ are the CO column density and the linear size of
the synthesized beam perpendicular to the jet axis. With $N \sim (4-8) \times 10^{20}$
\cms{} (see \S \ref{sec:TCD}), $b \sim 5\times10^{15}$ cm,
and $v_j \sim 100-200$ \vkm{}, $\dot{M}_j \sim (0.7-2.8)\times 10^{-6}$
\solarmass{} yr$^{-1}$, similar to that found in HH 212.
The accretion rate can be estimated assuming that the bolometric
luminosity $\Lbol$ is mainly from the
accretion. Assuming a stellar mass of $M_\ast \sim 0.06$ \solarmass{}
\cite[derived from an evolution model,][]{Froebrich2003}
and a stellar radius of $R_\ast \sim 4 R_\odot$ \citep{Stahler1980}, then 
the accretion rate 
$\dot{M}_a \sim \Lbol R_\ast/GM_\ast \sim 8 \times 10^{-6}$
\smassrate{}, with  $\Lbol \sim 3.6 L_\odot$ \citep{Froebrich2005}.
Thus, the mass-loss rate is estimated to be
$\sim 9-36$\% of the accretion rate, similar to
that found in HH 212 \citep{Lee2007}.

\subsection{Episodic, Bending, and Precessing} \label{sec:jetebp}

The jet is seen with a chain of knots in SiO and CO with a
semiperiodic spacing of $\sim$ \arcs{2}$-$\arcs{3} or 600$-$900 AU.
The knots may trace the unresolved (bow) shocks resulting from
a semiperiodic variation in the jet velocity \cite[see, e.g.,][]{Suttner1997}.
A temporal variation in the jet properties was also suggested in
\citet{Gueth1999}.
The period, which can be estimated by dividing 
the interknot spacing by the jet velocity,
is found to be $\sim 44-15$ yr, assuming a jet velocity of 100$-$200 \vkm{}.
The jet is believed to be launched from an accretion disk around the source.
The periodic velocity variation may be
due to (1)
periodic perturbation of the accretion disk by an unresolved companion
at a few AU away
or (2) magnetic cycle like the 
solar magnetic cycle, which has a period of $\sim$ 22 yr \citep{Shu1997}.

The jet seems to have a small-scale ($<$ 1\degree{}) precession
with a sinuous structure, in addition
to a large-scale ($\sim$ 3\degree{}) precession suggested in
\citet{Gueth1999} and \citet{Eisloffel2003}. 
Small-scale jet precession is
also seen in other jets, e.g., HH 34 and HH 212, and is generally ascribed
to the tidal effects of a companion star on the direction of the jet axis
\citep{Reipurth2002}. However,
it may also be due to kink instability in the jet \citep{Todo1993}. 

The jet is also seen bent in SiO with its eastern and western
components bent by $\sim$ 0.5\degree{} to the north.
The jet was also seen bent in CO but by 3 times larger 
\citep{Gueth1999} due to contamination from high-velocity
shells (see \S \ref{sec:COjs} \& \S \ref{sec:kinematics}).
The jet may have an additional bending into the plane of the sky because
the eastern and western components of the jet may have different 
inclinations due to their different mean velocities (see \S \ref{sec:dis_inc}).
Assuming their original paths of motion are
antiparallel, the eastern and western components of the jet
are both bent by $\sim$ 0.7\degree{}$-$1.4\degree{} into the plane of
the sky. Thus, the jet may have a total bending of $\sim$ 
0.9\degree{}$-$1.5\degree{}. 
Possible mechanisms for jet bending have been discussed in
\citet{Fendt1998} and \citet{Gueth1999}.
One of the possibilities is 
due to motion of the jet source in a binary system.
The envelope, which is seen
elongated toward the southwest, may result from an interaction with a
companion in the southwest.

\subsection{Rotating and Launching Radius?}


The jet is expected to be rotating,
carrying away extra angular momentum from
accretion disk. The HH 211 jet, being close to the plane of the sky, is one
of the best candidates to study the rotation. Here, only the first four
bright SiO knots (BK1, BK3, RK1, and RK2) are used, because the SiO knots
further out are likely to be more affected by shock interactions.

A velocity gradient is seen across the first four knots (see
the solid lines in Fig. \ref{fig:pvrotation}), with $\sim$ 1.5$\pm$0.8
\vkm{} at $\sim$ 30$\pm$15 AU (i.e., \arcsa{0}{1}) away from the jet axis.
The jet may have a radius (i.e., jet
edge) of $\sim$ 30 AU, similar to
that of HH 212, which is estimated to be $\sim$ 45 AU 
\citep{Cabrit2007}.
However, the beam dilution due to insufficient angular 
resolution could make the velocity shift smaller than it really
is \cite[see, e.g.,][]{Pesenti2004}.
This velocity gradient has the same direction as that seen in the rotating 
ammonia envelope (Wiseman, private communication), 
suggesting that it is from the jet rotation.
The velocity beyond $\pm$1.5 \vkm{} 
seems increasing toward the jet axis, also as predicted in
magneto-centrifugal wind models.

However,
the interpretation of the velocity gradient perpendicular to
the jet axis is complicated by the presence of a velocity
gradient along the jet axis (Fig. \ref{fig:maprotation}).
On the blueshifted (east) side, the higher-blueshifted emission (blue
contours) is upstream of the lower-blueshifted emission (red contours),
while on the redshifted (west) side,
the higher-redshifted emission (red contours) 
is upstream of the lower-redshifted emission (blue contours).
This velocity gradient along the jet axis is consistent with 
the SiO knots being formed by a velocity variation in the jet (see \S
\ref{sec:jetebp}), with the faster jet material catching up with the
slower jet material. Without sufficient angular resolution, 
this velocity gradient can not be separated from the 
component perpendicular to the jet axis.
The velocity gradient seen across the knots RK1 and BK1
may even arise from the velocity
gradient along the jet axis due to the beam position angle.
The jet also has a small-scale precession that may introduce 
a velocity asymmetry between the two jet edges 
\cite[see][]{Cerqueira2006}. 
The increase in velocity toward the jet axis may
also be due to the side-ways ejection of the shocked material.
Therefore, further observations at higher angular resolution are really needed
to confirm the jet rotation.

The launching radius of the jet can be estimated if
the measured velocity gradient is indeed from the jet rotation.
In magneto-centrifugal wind models, the jet can be considered as the dense
part of the wind along the rotational axis. In the wind,
the specific angular momentum and energy can be assumed to be conserved along
any given field line. Thus, for a given stellar mass, the specific
angular momentum and poloidal velocity (i.e., jet velocity)
at large distance
can be used to derive the angular velocity and thus the
wind launching radius at the foot
point of the field line in the accretion disk
\cite[see, e.g.,][]{Anderson2003}.
The angular velocity at the foot point is found to be 
$\Omega_0 \sim (0.82-3.23)\times10^{-6}$ s$^{-1}$ or 0.07$-$0.28 day$^{-1}$,
using Eq. 4 in \citet{Anderson2003}, assuming a stellar mass of $M_\ast \sim
0.06 M_\odot$ \citep{Froebrich2003} and a jet velocity of 100$-$200 \vkm{}.
Thus, the wind launching radius is
$\bar{\omega}_0 \sim 0.15-0.06$ AU or  $\sim (8-3) R_\ast$, with
the stellar radius $R_\ast\sim 4 R_\odot$ \citep{Stahler1980}.
Therefore, the jet could be launched either
from near the corotation radius ($\sim 0.05$ AU)
as in the X-wind model \citep{Shu2000} 
or beyond in the inner edge ($\sim 0.1$ AU) of a disk as in disk-wind
model \citep{Konigl2000}, depending on the jet velocity.
Note that, if the jet is really launched from near the corotation
radius, the rotation period of the protostar would be locked to be $\sim$ 22 days.
A classical T Tauri star has been found to have a typical rotation period of
$\sim$ 8 days \cite[see, e.g.,][]{Bouvier1993}, with a stellar mass of 0.5$-$0.8
\solarmass{}. The protostar here in HH 211, with a mass of $\sim$ 10 times smaller,
could have a rotation period of $\sim$ 3 times longer.

\section{Conclusions}
We have mapped the protostellar jet HH 211 in 342 GHz continuum, SiO
($J=8-7$), and CO ($J=3-2$) emission.  Thermal dust emission is seen in
continuum at the center of the jet, tracing an envelope and a possible
optically thick compact disk (with a size $<$ 130 AU) around the protostar. 
A knotty jet is seen in CO and SiO as in
\H2{}, but extending closer to the protostar. It consists of a chain of knots
on each side of the protostar, with an interknot spacing of $\sim$
\arcs{2}$-$\arcs{3} or 600$-$900 AU and
the innermost pair of knots at only $\sim$ \arcsa{1}{7} or 535 AU from the
protostar. These knots likely trace unresolved internal (bow) shocks in the jet, with a
velocity range up to $\sim$ 25 \vkm{}. The jet is episodic, precessing,
and bending. The two-sided mass-loss rate of the jet is
estimated to be $\sim (0.7-2.4)\times 10^{-6}$ \solarmass{} yr$^{-1}$, about
9$-$36\% of the accretion rate. 
A velocity gradient is seen consistently across 
two bright SiO knots (BK3 and RK2)
perpendicular to the jet axis, 
with $\sim$ 1.5$\pm$0.8 \vkm{} at $\sim$ 30$\pm$15 AU, suggesting a presence
of a jet rotation. 
The launching radius of the jet, derived
from the potential jet rotation, is $\sim$ 0.15$-$0.062 AU in the inner disk.

\acknowledgements
We thank the SMA staff for their efforts
in running and maintaining the array, and the anonymous referee for
the insightful comments. C.-F. Lee thanks Frank H. Shu,
Zhi-Yun Lee, and Nagayoshi Ohashi for fruitful conversations.
A. P. is grateful to Robert Estalella for helpful discussions.


\begin{figure} [!hbp]
\centering
\putfig{0.7}{270}{f7.ps}
\figcaption[]
{CO emission (0.3 to 20.3 \vkm{})
associated with the linear velocity structure in the west, 
which is marked with an ellipse in Figure \ref{fig:pvjet}, 
plotted on top of the IR image. The image is rotated by 26.6\degree{}
clockwise.
Contour spacing is 1.4 \Jybk{} (13.24 \Tk) with the first contour at 1.4 \Jybk{}.
The cross marks the source position. The Xs mark the positions of three SiO
knots.
The dashed line indicates the jet axis.
The dotted lines indicate the locations of the PV cuts shown in Figure
\ref{fig:pv_per}.
\label{fig:shell2}
}
\end{figure}

\begin{figure} [!hbp]
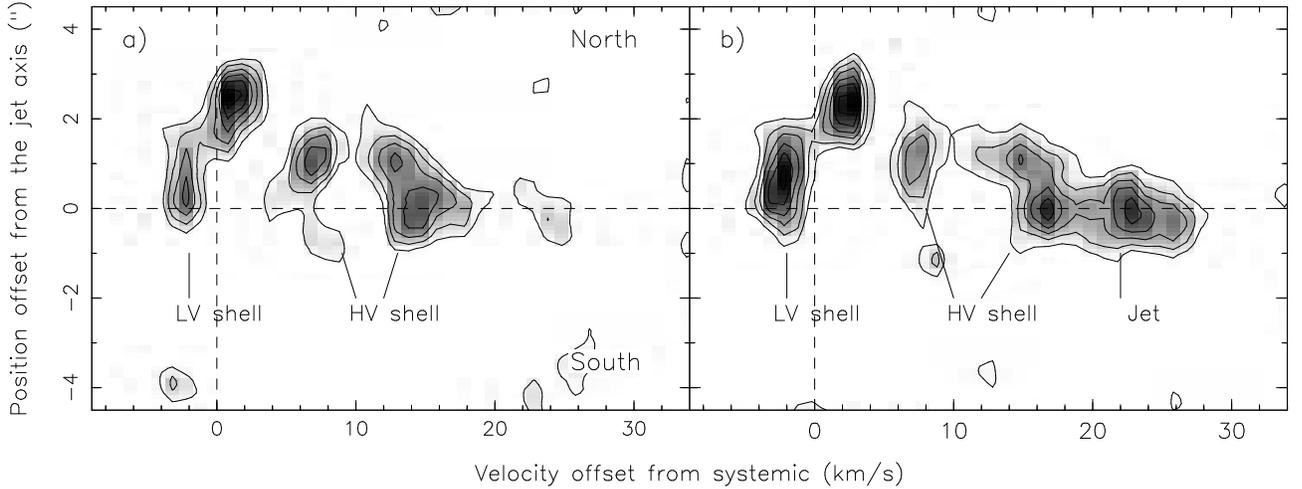

\centering
\putfig{0.7}{270}{f8.ps}
\figcaption[]
{CO PV cuts perpendicular to the jet axis in the redshifted lobe at 
\tlabel{a} \arcs{11}
and \tlabel{b} \arcs{12} away from the source. The cuts have a width of
\arcs{1}.
Contour spacing is 0.15 \Jyb{} (1.42 K) 
with the first contour at 0.2 \Jyb{} (1.89 K).
Here LV and HV shells mean the low-velocity and high-velocity shells, respectively.
\label{fig:pv_per}
}
\end{figure}

\begin{figure} [!hbp]
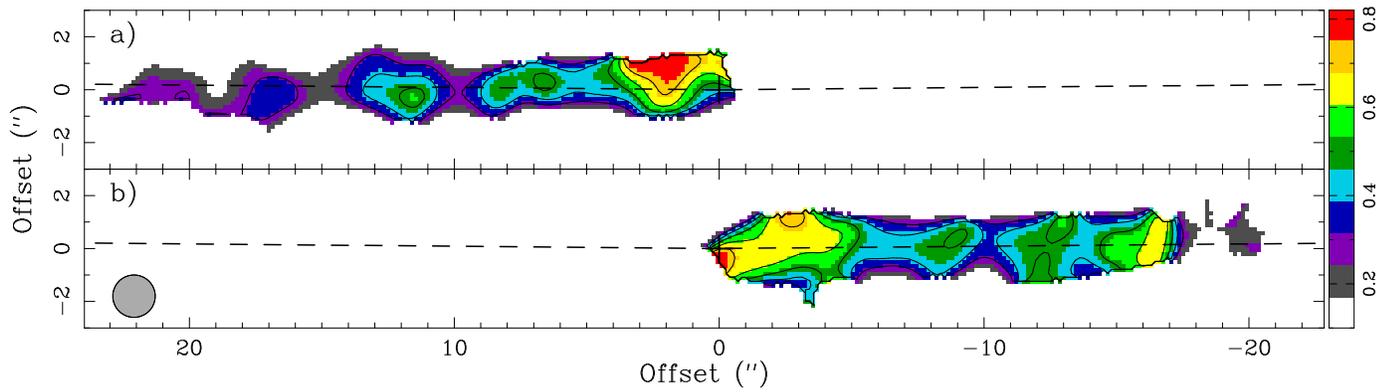

\centering
\putfig{0.8}{270}{f9.ps}
\figcaption[]
{Maps of SiO($J=8-7$)/SiO($J=5-4$) ratio, obtained from the
(integrated) SiO ($J=8-7$) and SiO($J=5-4$) maps that have been smoothed to
the same angular resolution of
\arcsa{1}{6}. \tlabel{a} Blueshifted side, with the emission
integrated from -22 to -2 \vkm{}.
\tlabel{b} Redshifted side, with the emission integrated from 6 to 34 \vkm{}.
Contour spacing is 0.1 with the first contour at 0.3.
\label{fig:SiOratio}
}
\end{figure}

\begin{figure}[!hbp]
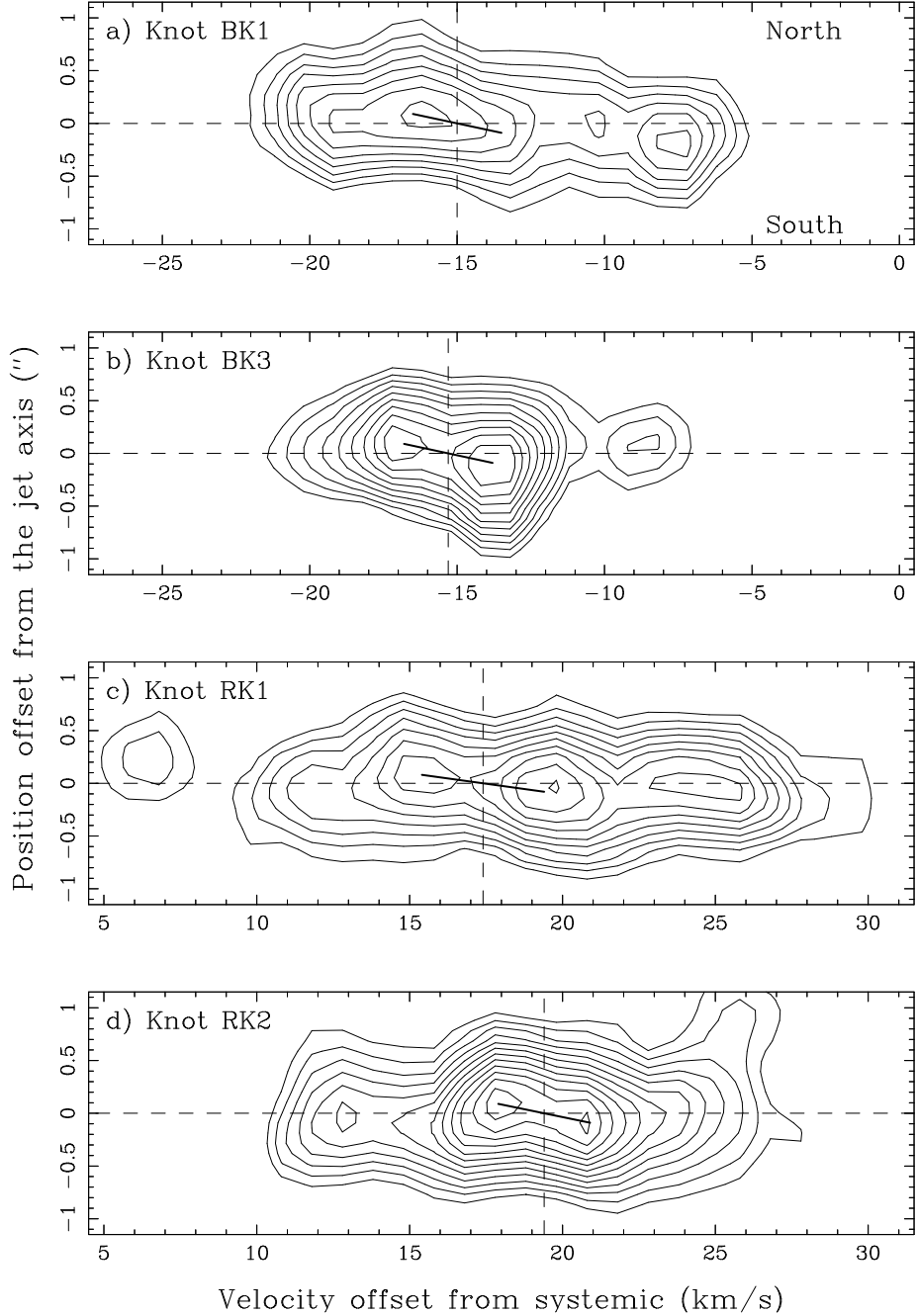

\centering
\putfig{0.7}{0}{f10.ps}
\figcaption[]{
PV diagrams of the SiO emission cut perpendicular to the jet axis
across the emission peaks of knots 
\tlabel{a} BK1, \tlabel{b} BK3, \tlabel{c} RK1, and \tlabel{d} RK2.
The cuts have a width of \arcsa{0}{3}. The solid lines define the velocity
gradients across the jet axis by connecting the two peaks on the opposite
sides. Contour spacing is 0.15 \Jyb{} (1.42 K)
with the first contour at 0.45 \Jyb{} (4.25 K).
\label{fig:pvrotation}
}
\end{figure}

\begin{figure}[!hbp]
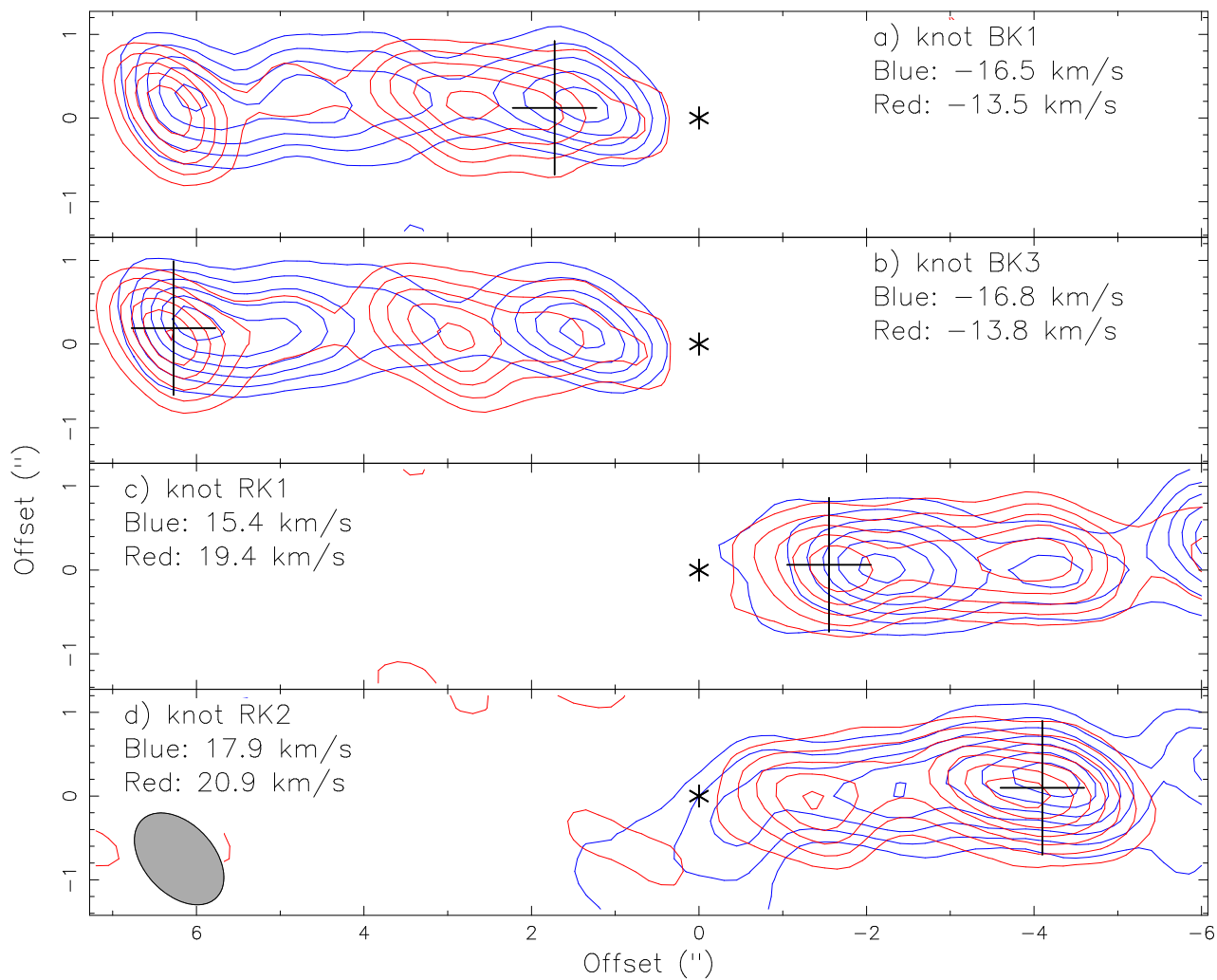

\centering
\putfig{0.8}{270}{f11.ps}
\figcaption[]{
1 \vkm{} wide maps of the two line peaks
that define their velocity gradient shown in Figure \ref{fig:pvrotation}
for knots \tlabel{a} BK1, \tlabel{b} BK3, \tlabel{c} RK1, and \tlabel{d} RK2.
Contour spacing is 0.45 \Jyb{} (4.25 K)
with the first contour at 0.30 \Jyb{} (2.83 K).
The crosses mark the cut centers where the emission peaks are in the
integrated maps (see Fig. \ref{fig:jet}c). 
The asterisks mark the source position.
\label{fig:maprotation}
}
\end{figure}

\end{document}